\PassOptionsToPackage{hyphens}{url}
\documentclass{czipreprint}

\usepackage{listings}
\usepackage{alltt}
\usepackage{dialogue}
\title{Use of LLMs in preparing accessible scientific papers}
\author[1]{Allison Doami}
\email{adoami@chanzuckerberg.com}
\author[1]{Christine James}
\email{cjames@chanzuckerberg.com}
\author[1]{Dan Lu}
\email{dlu@chanzuckerberg.com}
\author[1]{Lia Prins}
\email{lprins@chanzuckerberg.com}
\author[1]{Annette Torrence}
\email{atorrence@chanzuckerberg.com}
\author*[1,2]{Boris Veytsman}
\orcid{0000-0003-4674-8113}
\email{bveytsma@gmu.edu}
\affiliation[1]{Chan Zuckerberg Initiative, Redwood City, CA, USA}
\affiliation[2]{School of Systems Biology, George Mason University,
  Fairfax, VA, USA}

\begin{document}

\maketitle

\begin{abstract}
  Making scientific papers accessible may require reprocessing old
  papers to create output compliant with accessibility standards.  An
  important step there is to convert the visual formatting to the
  logical one.  In this report we describe our attempt at zero shot
  conversion of arXiv papers.  Our results are mixed: while it is
  possible to do conversion, the reliability is not too good.  We
  discuss alternative approaches to this problem.  
\end{abstract}

\section{Introduction}

Science becoming a mass occupation has been a revolutionary
transformation for the society.  The further acceleration of
scientific progress might be achieved by the tapping of the pool of
talents from the previously underrepresented populations: women,
minorities, disabled people.  Thus making science accessible for all
is not just the matter of equity.  It is important for all of us.

This task has both social and technical dimensions.  It is not enough
just to attract these populations to science.  To make contributions
of the disabled people possible, we need to solve the technical
problems of making work in science available for them.  The importance
of this task is being more and more recognized~\cite{ITU2024,
  OSTP2024}.

This is especially true with respect to people with low or no vision.
Scientific output consists of mostly papers and books: textual
information with images, tables etc.  We need to make them available
for assistive technologies like screen readers, Braille terminals,
etc.  There are standards for accessible Web pages~\cite{WCAG20,
  EN301, ICT508, WCAG21} and PDFs~\cite{PDFUA1, PDFUA2}.  An important
part of these standards is the preservation of the logical formatting
in the papers.  A printed page uses a number of visual cues to convey
the logical information to the reader.  A sighted person knows that a
bold-faced line probably is a section header, a centered line in the
beginning of a paper is probably a title, a line under it probably has
the information about authors, etc.  All these devices are described
as \emph{visual formatting.}  As the term itself suggests, this
formatting is intended for visual consumption.  It is more difficult
for assistive devices like screen readers to express the meaning for
the users.  The recommendations of EBU~\cite[\S~4.1.2]{EBU2024} insist
on the logical formatting:
\begin{quotation}
  Do not create manually individual headings by making them bold and
  by increasing the font size. Instead, create a style with this
  layout and apply it to every heading in your document. This will
  allow blind and partially sighted people to navigate easily through
  the document, using the heading list, generated by the screen
  reader. The advantages for you as an author are that it is easy to
  modify the style instead of all individual headings and that styles
  allow the automatic updating of your table of contents. 
\end{quotation}

Incidentally, logical formatting as opposed to visual formatting helps not
just the people with low vision, but also another category of non-sighted
consumers of texts: machines.  Logically formatted papers are much
more ``machine-readable'' and allow better analysis by various
automatic tools, which is important for classification and
summarization of knowledge.  This is an example of the well known
``curb-cut effect'': the changes made for the benefit of disabled
people often have unintended positive effects for the society in
general~\cite{Blackwell2016}.

\LaTeX\ is the system of choice of typesetting of papers in the
\acronym{STEM} fields.  Thus, the tendency of \LaTeX\ to encourage
logical formatting~\cite{Lamport94} is beneficial.  Unfortunately,
since \LaTeX\ was designed in the previous century when computers were
much less powerful than now, it is optimized to drop the information
about the logical structure very early in the typesetting process, so
the resulting \acronym{PDF} has only the visual information.  As the
author of the system underlying \LaTeX\ said as early as 1974,
``premature optimization is the root of all evil (or at least most of
it) in programming.''~\cite{Knuth:1974:CPA}.  The current effort of
\LaTeX3 project is to override this early decision and make \LaTeX\ to
produce \emph{tagged \acronym{PDF},} satisfying the accessibility
standards~\cite{Mittelbach:2024:LTP, Mittelbach:DocEng2024}.  Tagged
\acronym{PDF} files look like the traditional ones, but under the hood
they contain the additional information available to the specialized
software like screen readers.  This information enables the software
to discern the logical meaning of each element of the page, thus
making the reading more understandable.  Many publishers have the
\LaTeX\ source of the papers stored, so it will be possible to
reprocess the old papers to get the new accessible versions.

One of the biggest depositories of \acronym{STEM} papers in \LaTeX\
format is the arXiv preprint server that has about 2.4~million articles
and continues to grow fast.  The vast majority of arXiv papers are in
(La)\TeX\ format, and arXiv pledged to reprocess them into
accessible \acronym{PDF}~\cite{Preining:2024:TLA} format.

In preparation for this step, we need to assess whether the current
papers in arXiv are suitable for conversion.  While \LaTeX\ encourages
logical formatting, it does not enforce it, and some, especially older
papers, indeed use visual instead of logical formatting.  Spot checks
show that this is the case for 5\%--10\% of preprints, which
translates to hundreds thousands of papers that need to be addressed.
Most papers with visual formatting are rather old (before 2020), but
even some new papers use it.

The translation from visual to logical formatting, especially at
scale, is not trivial.  If we had known that authors, for example,
always use bold large fonts for section headers, we would write a
simple regular expression to change
\verb|\textbf{\large 1 Introduction}| in a solitary paragraph to
\verb|\section{Introduction}|.  Unfortunately,authors are quite
inventive. There are too many ways to set up visual formatting---and
almost any of them is used by an author!

This looks like an interesting problem to try machine learning
methods.  Can an algorithm convert the visual formatting to logical
formatting?  While \acronym{LLM}s are ultimately language processing
machines rather than information processing ones~\cite{Walters2023},
the conversion from visual format to a logical format can be seen as
ultimately a language task, so hopefully \acronym{LLM} can deal with
it.  

In September 2024, Chan Zuckerberg Initiative organized a hackathon for
its employees.  We decided this problem might be a good idea for a
hackathon: an exploratory research by a dedicated team.

We downloaded a sample of papers from arXiv, and selected those with
elements of visual formatting.  We interactively talked to ChatGPT~4o
from OpenAI~\cite{openai2023chatgpt} and compared the results with the
expected ones.  

While this research was completely exploratory, we decided to publish
the results of our three days intense conversations with ChatGPT to
show both the promises and pitfalls of this approach.

\section{Dialogs with ChatGPT about \LaTeX{}}
\label{sec:dialogs}

\subsection{The first attempt: assume you are a \LaTeX\ expert\ldots}
\label{sec:1st_attempt}

One of the most exciting features of the current iteration of AI is
its ability to write and rewrite code.  Having read many programs,
modern LLMs are quite impressive in generating new ones.  Today many
programmers, especially young ones, use different versions of LLMs to
write (at least the initial versions) of their code.

We initially assumed that having read source code of many papers
(while the training corpus for most LLMs remains closed, it is
reasonable to assume it includes many scientific texts, for example,
from arXiv), ChatGPT became a true expert in \LaTeX.

This led to the following prompt:
\begin{dialogue}
  \speak{Prompt} You are a LaTeX expert.  Your task is to make LaTeX
  files readable to blind people.  Blind people can understand logical
  formatting like ``\verb|\section{Introduction}|''.  However, they have
  problems with visual formatting, like
  ``\verb|\textbf{\Large Introduction}|''.  You want to change all
  instances of visual formatting to the logical formatting.  This
  should include converting ``\verb|{\bf Definition}|'' to theorem
  environment.  Please convert the file.
\end{dialogue}

Unfortunately, this did not work.  ChatGPT was overzealous in changing
boldface and italics to section headers, and did not take care of the
context.  This led to disastrous results where content meaning was changed. For example,sometimes this formatting is used in mathematical formulas, like a vector \verb|${\bf v}$| (of course, the recommended way to typeset a vector would be
\verb|$\mathbf{v}$|, which provides a clue about the meaning, but an
author using visual formatting for sections does not usually pay attention to these details). In this example, the formula syntax is broken by ChatGPT's attempt to make it more accessible.

\subsection{The second attempt: more clues\ldots}
\label{sec:clues}

We tried to be more explicit with the next prompt, adding the following
caveats: 
\begin{dialogue}
  \speak{Prompt} Sometimes bold text means something else than section
  heading though. For example, the case
  \verb|\textbf{\bibinfo{volume}{3}}| means ``make volume number
  bold''. Please reprocess the file.

  \speak{Prompt} Changing bf to textbf and it to textit is not
  helpful, etc. I want you to evaluate what kind of words the paper is
  using to bold font and determine if it is a section or just an
  emphasized word. When you come across bf or it and it is not a
  header, then convert it to emph. 
\end{dialogue}

However, this prompt engineering was rather fragile.

At this point we decided to experiment with the metadata discovery,
assuming that extraction of paper's authors, title and affiliations
would be an easier task.

\subsection{The third attempt: let's try metadata and external knowledge}
\label{sec:metadata}

While guessing paper logical structure may be a very difficult task,
the determination of paper metadata might be a simpler one.  By
metadata we mean title, authors, affiliations and abstract.  Visual
formatting of these entities may include centered text, large font,
bold font, and---for affiliations---markers like $\ast$, $\dagger$,
etc.

We also added some knowledge base to the things ChatGPT considers,
giving it examples of logically and visually structured files.  Here
is a typical dialog with ChatGPT:

\begin{dialogue}
  \speak{Prompt} You are a LaTeX expert. Your task is to make LaTeX
  files readable to blind people.  Blind people would use a screen
  reader to read papers.  Inaccessible LaTeX files will use visual
  formatting (text properties) to indicate properties of a document
  that would be specifically identified using specific key words in
  accessible LaTeX. For example, an inaccessible LaTeX file may use
  bold, italics, centering, whitespace, and other text properties to
  format an author's name. An accessible LaTeX file would simply use
  the author command. An inaccessible LaTeX file may use visual
  formatting to make a document title more prominent. An accessible
  LaTeX file will use the title command or the maketitle command.  Can
  you make this file accessible?

  \speak{ChatGPT} The LaTeX file uses a lot of custom formatting that
  relies on visual properties such as bold, italics, and centering
  rather than semantic commands. To make it accessible for screen
  readers, I will replace these visual formatting commands with
  appropriate LaTeX commands such as \verb|\title|, \verb|\author|,
  \verb|\maketitle|, \verb|\section|, and \verb|\textit|. Here's an
  outline of the necessary changes:

  Title: Replace \verb|\centerline{\bf ...}| with \verb|\title{...}|.
  
Author: Replace the visual formatting of the author name (currently
\verb|\centerline{Giuseppe Gaeta}|) with the \verb|\author{...}| command.

Institution and Email: Replace the current formatting for the
institution and email with the appropriate \verb|\affil{...}| command (or
include it as part of the \verb|\author| information).

Abstract: Use the \verb|\begin{abstract} ... \end{abstract}| environment
instead of \verb|{\bf Abstract. }| and \verb|{\it ...}| for the
abstract content. 

Sections: Replace custom section formatting (\verb|\section#1|) with
proper LaTeX section commands like \verb|\section{...}|.

I'll proceed to modify the file accordingly.

The LaTeX file has been modified to be more accessible. You can
download the accessible version of the file from the following link: 

Download Accessible LaTeX File

This version replaces visual formatting with semantic LaTeX commands
suitable for screen readers.
\end{dialogue}

This example highlights the importance of data integrity within the body of the file. As you can see, ChatGPT chose to update sections (ineffectively) without specific instructions to do so. We included explicit instructions to avoid editing the body of the paper in later iterations of the prompt, which was usually effective, but not reliably so.

We also tried a simplistic approach where we supplied a paper that manually verified to be accessible and asked ChatGPT to update another file to be similarly accessible. In this case, ChatGPT replaced the author of the paper formatted with visual formatting with the author from the accessibly formatted paper. Additional training data and more explicit prompts might help this type of approach be feasible in future efforts.

In these and numerous other cases, ChatGPT tended to create non-compilable files, with senseless commands like
\begin{displaymath}
\verb|\def\giorno{15/6/98\end{abstract}|.    
\end{displaymath}
Only after several rounds of dialog ChatGPT was able to produce satisfactory result.

\section{Discussion}
\label{sec:discussion}

In our experiments zero-shot transformation of text was possible.  
A good picture of the challenges faced can be inferred from a
single ChatGPT prompt we created
(see~Appendix~\ref{sec:prompt}).  Each phrase of this prompt was added
with intention to address specific issues encountered during
experimentation on different papers with variable visual formatting. 

The ways of visual formatting of the elements were quite
different and numerous: authors employed font changes, size changes, centering,
etc.  The most difficult task is the parsing of affiliations: authors
often have several affiliations, which are mapped to the authors in a
``many-to-many'' relation.  In some cases, the affiliations may be expressed
in tables, inserted within the front matter text, or, more commonly, expressed
with footnote-like markers, which are quite different from paper to paper.

As a result, the transformation was not reliable enough to get good
results without human intervention and supervision.  In one experiment,
a single ChatGPT prompt was able to successfully transform 7 of 10 visually
formatted papers.  In another example, the LLM changed the authors of the paper
to make it more “accessible” by transferring an author from a logically formatted
paper given to ChatGPT as an example into the transformed paper.  This shows
that ChatGPT has difficulty distinguishing between formatting and content.  
Note that even when we asked ChatGPT to change only the metadata,
it still occasionally changed the body of the article.  At times during our
experiments, ChatGPT would ignore instructions in the prompt or,
more dangerously, state in the prompt response that it followed the instructions
when it had actually inappropriately made those edits to the transformed file.  
In many cases ChatGPT created non-compilable papers that generated \LaTeX errors.

It should also be noted that ChatGPT was able to learn how to create compilable 
and logically formatted \LaTeX when provided with multiple, iterative instructive 
prompts.  It's possible that a custom ChatGPT bot trained on a large quantity of 
representative data might be able to overcome some of the limitations encountered 
during these experiments.

For bulk conversion the biggest task is data validation, especially
since there could be many edge cases.  Validation of any transformed
text is extremely important due to the generative nature of AI
technology that powers ChatGPT and other LLMs.  In addition to ``honest
mistakes'' when identifying and transforming visual to logical
formatting (e.g. incorrectly mapping an author to another author's
affiliation), generative AI may ``hallucinate'' and insert completely
unrelated content anywhere within the paper.  Previous work shows that
hallucinations are very common when generating bibliographic citations
for \acronym{LLM}-generated academic papers referencing real academic papers.  A
significant percentages of citations generated by ChatGPT in these
experiments were completely fabricated~\cite{Walters2023}, which is
not surprising since \acronym{LLM}s are language-processing rather than
information processing tools.

This exploratory research had a number of
factors that helped us avoid complications that LLMs can present.
Focusing on metadata (material before
\verb|\section{Introduction}| [...] \verb|\maketitle|, or the end of
the abstract) allowed us to circumvent a limitation of the modern LLMs:
the short window for reading.  Another mitigating factor was the fact
that the authors usually have a mostly uniform order of metadata
presentation: a title is first, followed by the author(s), then
affiliations and an abstract with optional keywords.  A
systematic approach to the parsing of metadata may include the use of the
arXiv API that provides ground truth for titles, authors and
abstracts.  Note that our experience shows that the affiliation metadata in arXiv
API is less complete than the affiliation in the papers themselves.  

An alternative approach to the use of ChatGPT might be the deployment
of a specially trained model to recognize visual formatting and
convert it to logical formatting.  An approach to develop this model
might be based on the ideas discussed in~\cite{CitationParsing19,
  BibTeX2020}:
\begin{enumerate}
\item Take a large database of logically formatted papers.
\item Use a script to convert it to visually formatted manuscript.
\item Train an algorithm to convert it back.
\end{enumerate}

\section{Conclusions}
\label{sec:concl}

The zero-shot approach to convert visual to logical formatting gives
tantalizingly good results, which unfortunately are not acceptable
without human supervision.  A specially trained model or a custom GPT trained on a more extensive and diverse data set might be a better solution for the problem.

\bibliographystyle{tugboat}
\bibliography{main}
\makesignature

\clearpage
\appendix

\section{The final prompt}
\label{sec:prompt}

Below is the final prompt developed in our experiments

\begin{quotation}
  You are a latex expert. Your task is to make latex files readable to
  blind people.  Blind people would use a screen reader to read
  papers.  Inaccessible latex files will use visual formatting (text
  properties) to indicate logical properties of a document that would
  be specifically identified using specific key words in accessible
  latex. For example, an inaccessible latex file may use bold,
  italics, centering, whitespace, and other text properties to format
  a document title. To make this file more accessible, replace an
  inaccessible title with visual formatting and no title command with
  an accessible title command. The title in an inaccessible latex
  paper will often be formatted with center, bold, italics or similar
  and is near the beginning of the document. If the title is already
  inside a title command, it is ok to retain visual formatting like
  bold and italics. Some title commands include a short version in
  square brackets and a longer version in curly braces. This is
  accessible, but is not required to be accessible. Can you make this
  file accessible? Do not change anything in this file other than the
  title.  The author and affiliation are not part of the title. After
  you make changes, check the resulting file to make sure that it has
  been updated correctly. 

  Accessible latex files should use the ``\verb|\author{}|'' and
  ``\verb|\affiliati|\penalty0\verb|on{}|'' commands. After the title is correctly
  formatted, identify the authors (one or more names of people who
  wrote the paper) and their affiliations (the organizations, such as
  a university, where the authors are affiliated). These accessible
  latex commands, ``\verb|\author{}|'' and ``\verb|\affiliation{}|''
  should replace visual formatting such as bold, italics, centering,
  whitespace, and other text properties to format the authors and
  their affiliations. In an accessible file, each author or
  affiliation should be specified with a separate author or
  affiliation command. A ``\verb|\thanks|'' command is also an
  accessible way to express affiliation, though not preferred. Each
  affiliation should be placed immediately below the affiliated author
  and should replace any previous text used to indicate
  affiliations. Affiliations can span multiple lines in an
  inaccessible latex file and should be combined into a single command
  affiliation to be accessible. Sometimes authors and affiliations may
  have no visual markup at all. In those cases, they are usually in
  close proximity to the title. Look for symbols or numbers that map
  authors to affiliations as a clue.

Some author commands include a short version in square brackets and a longer version in curly braces. This is accessible, but is not required to be accessible. It's also acceptable to include visual formatting inside names, since sometimes people have special characters like ``\verb|\H{o}|'' and ``\verb|\'{a}|'' inside their name.

Subscript or superscript may be helpful in cases with multiple authors
and multiple affiliations. These subscripts or superscripts should be
removed after using them to find the correct affiliations for the
authors. After you make these changes, check the names of the authors
to make sure that they do not include any subscript or superscript
text or symbols. 

Accessible latex files should also have a maketitle command placed
after all front matter. The maketitle should never include extra
characters such as ``\verb|}|''. Make sure that the maketitle is placed after
the title, affiliation, author and similar commands nearby. Sometimes
a latex file may include a def command that contains a maketitle
command. If it does, then this new defined version of maketitle can be
used in accessible files instead of the normal maketitle. The
maketitle command is usually placed immediately after the front
matter, but may be placed later in the file before the first
section. After you make changes, check the resulting file to make sure
that it has been updated correctly. 

Identify the abstract within the latex file. Remember that an abstract
is one or more paragraphs that describe the paper and usually is
placed right after the author and affiliation information and before
any sections that might exist. It may also be inside a titlepage and
would end when the titlepage ends or when a new section begins. An
accessible latex file will place the abstract between begin and end
abstract commands. Inaccessible files may use visual formatting to
label the abstract. The abstract of an accessible paper will be placed
below the rest of the front matter.

\end{quotation}

\end{document}